\documentclass[12pt]{article}
\usepackage{amssymb,amsmath,slashed,latexsym}
\usepackage{float}
\allowdisplaybreaks
\newcommand{\ft}[2]{{\textstyle\frac{#1}{#2}}}

\newcommand{\bbox}{\lower.2ex\hbox{$\Box$}}
\newsavebox{\uuunit}
\sbox{\uuunit}
    {\setlength{\unitlength}{0.825em}
     \begin{picture}(0.6,0.7)
        \thinlines
        \put(0,0){\line(1,0){0.5}}
        \put(0.15,0){\line(0,1){0.7}}
        \put(0.35,0){\line(0,1){0.8}}
       \multiput(0.3,0.8)(-0.04,-0.02){12}{\rule{0.5pt}{0.5pt}}
     \end {picture}}

\csname @addtoreset\endcsname{equation}{section}
\newcommand{\SU}{\mathop{\rm SU}}

\newcommand{\U}{\mathop{\rm {}U}}

   \usepackage[nosort]{cite}
   \pdfoutput=1
  \usepackage[pdftex]{hyperref}
\usepackage{color}
\newcommand{\dr}{\raise.3ex\hbox{$\stackrel{\leftarrow}{\delta  }$}{}}
\newcommand{\dl}{\raise.3ex\hbox{$\stackrel{\rightarrow}{\delta }$}{} }
\newcommand{\pl}{\raise.3ex\hbox{$\stackrel{\rightarrow}{\partial }$}{} }

\begin{document}

 %%%%%%%%%%%%%%%%%%%%%%%%%%%%%%%%%%%%%%%%%%%%%%%%%%%%%%%%%%%
\begin{titlepage}
\begin{flushright}
CERN-TH-2017-189
\end{flushright}
%\phantom{.}
\vspace{.5cm}
\begin{center}
\baselineskip=16pt
{\LARGE  Highlights in Supergravity: CCJ 47 Years Later}\\
\vfill%\vskip 15mm%27.mm
{\large  {\bf Sergio Ferrara}$^{1,2,3}$ {\bf  and Marine Samsonyan}$^1$ } \\
\vfill

{\small$^1$ Theoretical Physics Department, CERN CH-1211 Geneva 23, Switzerland\\\smallskip
$^2$ INFN - Laboratori Nazionali di Frascati Via Enrico Fermi 40, I-00044 Frascati, Italy\\\smallskip
$^3$ Department of Physics and Astronomy and Mani L.Bhaumik Institute for Theoretical Physics, U.C.L.A., Los Angeles CA 90095-1547, USA\\}
\end{center}
\vfill
\begin{center}
{\bf Abstract}
\end{center}
{\small 
We consider an expression for the supercurrent in the superconformal formulation of $N=1$ supergravity. A chiral compensator provides the supersymmetric formulation of the Callan-Coleman-Jackiw (CCJ) improved stress energy tensor, when the conformal gauge is used. Superconformal and non-superconformal matter give different conservation laws of the supercurrent, when coupled to the curvature supermultiplets which underlie the local superspace geometry. This approach can be applied to any set of auxiliary fields and it is useful to classify rigid curved superspace geometries. Examples with four supersymmetries are briefly described.}
\vfill

\begin{center}
{\it Contribution to the Proceedings of the Erice International School of Subnuclear Physics, \\55th Course: 
``Highlights from LHC and the other frontiers of physics"\\ Erice, 14-23 June 2017}
\end{center}

\end{titlepage}

\addtocounter{page}{1}
 \tableofcontents{}
\newpage
%%%%%%%%%%%%%%%%

\section{Introduction: 
 Resurrection of the Energy Momentum Tensor}

In 1970, during the 8th Course of the Erice International School of Subnuclear Physics: ``Elementary Processes at High Energies", Sidney Coleman gave one of his celebrated lectures on a  ``New Energy Momentum Tensor" following a paper by Callan, Coleman, Jackiw (CCJ)  \cite{Callan:1970ze} . 
This paper solved the problem of defining an improved stress tensor $\Theta_{\mu\nu}$, different from the canonical one. This tensor is conserved and symmetric.
\begin{eqnarray}
\partial^\mu \Theta_{\mu\nu}= 0\, , \qquad
\Theta_{\mu\nu}=\Theta_{\nu\mu}\, ,
\end{eqnarray}
as a consequence of Poincar\`e invariance in flat Minkowski space, but moreover it is also traceless
\begin{equation}
\Theta_\mu^{\,\,\mu}=0\, ,
\end{equation}
when the theory is scale and conformal invariant.  In this way the non vanishing of the trace is an operator which measures the departure from scale and conformal symmetry. This breaking can occur at the classical level due to dimensionful terms in the scalar potential $V(\phi)$
 \begin{equation}
 \Theta_\mu^{\,\,\,\mu}= -4V+ V_{, i} \phi^i\, ,
 \end{equation}
or it can, be due to quantum effects either perturbative (non-vanishing $\beta(g)$ function in perturbation theory) or non-perturbative, such as instanton effects in supersymmetric gauge theories.  It is the aim of these lectures to discuss the energy-momentum tensor \cite{Ferrara:2017yhz} in supersymmetric field theories and its role in the breaking of ``superconformal symmetry". When considered as a source of the gravitational fields in curved space, the supergravity couplings to the stress tensor multiplet defines the (Super)-Einstein equations, whose source is the ``Supercurrent", introduced in \cite{Ferrara:1974pz}.  We will use an off-shell formalism based on the superconformal formulation \cite{Ferrara:1978jt,Cremmer:1982en,Freedman:2012zz}   of local superspace geometry.   The content of this lectures is mainly based on the results of \cite{Ferrara:2017yhz}. 

\section{The Improved Stress Tensor, Conformal Algebra and its Noether Currents}
The Noether currents of the ``conformal algebra" can be written in a unified way in terms of the CCJ tensor
\begin{equation}
J_\mu^\xi=\Theta_{\mu\nu}(x) \xi^\nu(x) \, .
\end{equation}
The corresponding charges 
\begin{equation}
Q^\xi =\int d^3xJ_0^\xi \, ,
\end{equation} 
are conserved when $\partial^\mu J_\mu ^\xi =0$. This happens when $\partial^\mu\Theta_{\mu\nu}=0$,  $\Theta_\mu^{\,\, \mu}=0$, provided the displacement $\delta x^\mu =\xi^\mu (x)$ satisfies the following equation for $d>2$
\begin{equation}
\partial_\mu \xi_\nu (x)+\partial_\nu \xi_\mu (x)=\frac{2}{d}\eta_{\mu\nu} \partial^\lambda \xi_\lambda (x)\, . \label{diffeq}
\end{equation}
For $d>2$, in particular $d=4$, the solution of this equation is
\begin{align}
&\xi^\mu (x)= a^\mu + \omega ^{\mu\nu}x_\nu +\lambda x^\mu +2 x^\mu x\cdot c -x^2 c^\mu \, , \label{xmu}\\
& a^\mu,   \,\, \omega^{\mu\nu}=-\omega^{\nu\mu} \rightarrow \mbox{Poincar\`e algebra (translation and Lorentz rotations)}\, ,\label{aomegamu}\\
&\lambda, c_\mu \rightarrow \mbox{dilatation and special conformal (boosts) transformation}\, .\label{cmu}
\end{align}

This algebra has 15 generators  $P_\mu, M_{\mu\nu}, D, K_\mu$, satisfying the commutation relation of the Lie algebra $SO(4,2)$ (or $SU(2,2)$). 
\begin{eqnarray}
& [M_{\mu\nu},D]=0,  \quad [P_\mu, D]=i P_\mu,  \quad [K_\mu,D]=-i K_\mu, \quad [K_\mu, K_\nu]=0 \, ,\\
& [M_{\mu\nu}, K_\rho]=-i(g_{\rho\mu}K_\nu-g_{\rho\nu} K_\mu), \qquad   [P_\mu, K_\nu]=2i (g_{\mu\nu}D-M_{\mu\nu})\, .
\end{eqnarray}
Note that for $a^\mu, \omega^{\mu\nu}$ $\partial_{(\lambda}\xi_{\mu )}=0$ so the differential equation \eqref{diffeq} is empty.  When $\Theta_\mu^{\,\,\mu}\neq 0$ the $D, K_\mu$ charges are not conserved.

For finite transformations $a_\mu, \Lambda_\mu^{\,\,\, \nu}, e^\lambda , c_\mu$ we have the action of the conformal group
\begin{align}
&\mbox{Poincar\`e} \,\,\,\,\,\,\,\,\,\,\, x_\mu ^\prime =a_\mu +\Lambda_\mu^{\,\,\, \nu} x_\nu \quad (\Lambda_\mu^{\,\,\,\nu} \eta_{\nu\rho} \Lambda^\rho_{\,\,\, \sigma}=\eta_{\mu\sigma})\, , \label{Ponx}\\
&\mbox{Dilatation}\quad x_\mu^\prime =e^\lambda x_\mu \, ,\label{dilat}\\
&\mbox{Conf transf} \,\,\, x_\mu^\prime =\frac{x_\mu +c_\mu x^2}{1+ 2 c \cdot x +c^2 x^2} \,\,  
\left(x^{\prime 2}=\frac{x^2}{1+ 2c \cdot x+c^2x^2}, x^2=0  \Rightarrow x^{\prime 2}=0 \right)\, \label{xprime},
\end{align}
in the infinitesimal $\delta x^\mu =\xi^\mu (x)$. Equations \eqref{Ponx},\eqref{dilat},\eqref{xprime} reduce to \eqref{xmu}.

For generic dimension $d>4$, the conformal group is $SO(d,2)$ with the following generators:
\begin{eqnarray}
&&M_{\mu\nu} \quad SO(d-1,1)\, ,  \\
&& P_\mu,  K_\mu \quad  d\,\, \mbox{vectors}\, , \\
&& D\,\, \mbox{dilation (1 generator)}\, .
\end{eqnarray}
For certain systems the canonical and improved energy momentum tensors coincide. This is the case of electromagnetic  field, whose stress tensor obeys $\Theta_\mu^{em\,\, \mu}=0$ because of its
definition:
\begin{equation}
\Theta_{\mu\nu}^{em}=F_{\mu\rho} F_\nu^{\,\,\rho}-\ft 14 \eta_{\mu\nu} F_{\sigma\rho}F^{\sigma\rho} \quad \mbox{with} \quad F_{\mu\nu}(A)=\partial_\mu A_\nu-\partial_\nu A_\mu \, .
\end{equation} 
Moreover, by using the cyclic identity $\partial_\rho F_{\mu\nu}+\partial_\mu F_{\nu\rho}+\partial_\nu F_{\rho\mu}=0$, one can show that
\begin{equation}
\partial^\mu \Theta_{\mu\nu}^{em}=\left(\partial^\mu F_{\mu\rho}\right) F_\nu^{\,\, \rho}=J_\rho F_\nu^{\,\,\rho}\, ,
\end{equation}
where for pure electromagnetism, $J_\rho =0$.
If $J_\rho \neq 0$, this term is cancelled by the matter part, $\partial^\mu \Theta_{\mu\nu}^M$, since $\Theta_{\mu\nu}=\Theta_{\mu\nu}^{em}+\Theta_{\mu\nu}^M$.

The simplest theory where an improvement term occurs is the theory of one (or more) neutral scalar fields $\varphi^i$ with a potential $V(\varphi^i)$. With a mostly plus metric the Lagrangian for this system is given by
\begin{equation}
{\cal L}=-\ft 12 g^{\mu\nu}\partial_\mu \varphi^i \partial_\nu \varphi^i -V(\varphi^i)\, .
\end{equation}
The canonical energy momentum tensor for this system is
\begin{equation}
T_{\mu\nu}=g_{\mu\nu}{\cal L}+\partial_\mu \varphi^i \partial_\nu \varphi^i \, .
\end{equation}
It is conserved using the $\varphi$ equations of motions
\begin{equation}
\partial^\mu T_{\mu\nu}=\partial_\nu \varphi \left(\Box \varphi^i -\frac{\partial V}{\partial \varphi ^i}\right)=0\, .
\end{equation}
However, this tensor is not traceless
\begin{equation}
T_\mu^{\,\,\mu}=-\partial_\mu \varphi^i \partial^\mu \varphi ^i -4 V(\varphi^i)\neq 0 \, ,
\end{equation}
even when the equations of motions are used.
Let's then define a new stress tensor $\Theta_{\mu\nu}$
\begin{equation}
\Theta_{\mu\nu}=T_{\mu\nu}-\ft 16 \left(\partial_\mu \partial_\nu -g_{\mu\nu}\Box \right) \varphi^{i \, 2}\, ,
\end{equation}
which is obviously conserved $\partial^\mu \Theta_{\mu\nu}=0$. However, the second term now contributes to its trace
\begin{equation}
\Theta_\mu^{\,\,\mu}=T_\mu^{\,\, \mu}+\ft 12 \Box \varphi^{i\,2}= -4V +\varphi^i \frac{\partial V}{\partial \varphi^i}\, .\label{ThetaVmu}
\end{equation}
In order to derive eq. \eqref{ThetaVmu} we used the equations of motions. For a scale invariant theory $\Theta_\mu^{\,\,\mu}=0$ since $V(\varphi)$ is homogeneous of degree 4.

Let us now consider a $U(1)$ gauge theory of complex scalar fields $\varphi^i$. The scalar part of the Lagrangian is given by
\begin{equation}
{\cal L}^M=(\partial_\mu -i A_\mu)\varphi^i (\partial^\mu+i A^\mu)\bar \varphi ^i -V(\varphi, \bar\varphi)\, ,
\end{equation}
with the canonical energy momentum tensor
\begin{equation}
T_{\mu\nu}^M=g_{\mu\nu}{\cal L}^M+\left((\partial_\mu -i A_\mu)\varphi^i (\partial_\nu +i A_\nu)\bar \varphi ^i +(\mu \leftrightarrow \nu)\right) \, .
\end{equation}
To define a traceless energy-momentum tensor $\Theta_{\mu\nu}^M$, we add the improvement term  $-\ft 13 (\partial_\mu \partial_\nu -g_{\mu\nu}\Box)\varphi^i  \bar \varphi ^i$ to $T_{\mu\nu}$. It will be traceless thanks again to the fact that $V(\varphi, \bar \varphi)$ is homogeneous of degree 4 for scale invariant theories
\begin{equation}
\Theta_\mu ^{M\,\, \mu}=4V-V_i \varphi ^i -V_{\bar i} \bar \varphi ^i =0\, ,
\end{equation}
but its divergence is non vanishing 
\begin{equation}
\partial^\mu \Theta_{\mu\nu}^M=-\partial^\mu \Theta_{\mu\nu}^{em}=J^\mu F_{\mu\nu}(A)\neq 0\, ,
\end{equation}
where
 \begin{eqnarray}
&J_\mu (\varphi, A)=i \left(\varphi^i D_\mu \bar \varphi^i -D_\mu \varphi^i \bar \varphi^i \right)\, , \\
&D_\mu \varphi^i =(\partial_\mu -i A_\mu)\varphi^i \,  ,\qquad D_\mu \bar \varphi^i= (\partial_\mu +i A_\mu)\bar \varphi ^i \, .
\end{eqnarray}
In supergravity a $U(1)$ is the R-symmetry of the superconformal algebra (which is a gauge symmetry in the conformal part of the action.),
but a term $\Theta_{\mu\nu}^{em}$ is missing, since in Einstein supergravity the $A_\mu$ gauge field is non-propagating. It is actually an auxiliary field
 \cite{Ferrara:1978em, Stelle:1978ye}, which  appears quadratically in the pure supergravity part of the action
\begin{equation}
{\cal L}=\sqrt{-g}\kappa^{-2}\left(\ft 12 R +3 A_\mu^2+...\right)\, .
\end{equation}
However, $A_\mu$ contributes to the Einstein equations $\frac{\delta {\cal L}}{\delta g^{\mu\nu}}=0$ and indeed it contributes to the covariant divergence of the Einstein equations ($G_{\mu\nu}=R_{\mu\nu}-\ft 12 g_{\mu\nu} R$ is the Einstein tensor)
\begin{equation}
\frac{2}{\sqrt{-g}}\frac{\delta{\cal L}}{\delta g^{\mu\nu}}=G_{\mu\nu}+6 A_\mu A_\nu - 3 g_{\mu\nu} A_\rho A^\rho+...=0\, .
\end{equation}
So  the $A_\mu$ term contributes to both the trace and divergence of the Einstein equation since
\begin{eqnarray}
&\frac{2}{\sqrt{-g}}\frac{\delta {\cal L}}{\delta g^{\mu\nu}} g^{\mu\nu}=-R-6 A_\mu^2+...=0\, ,\\
&\nabla^\mu\left(\frac{2}{\sqrt{-g}}\frac{\delta {\cal L}}{\delta g^{\mu\nu}}\right)=6 \nabla^\mu A_\mu \cdot A_\nu + 6 A^\mu F_{\mu\nu}(A)+...=0\, .
\end{eqnarray} 
For superconformal matter it turns out that $R+ 6 A_\mu^2=0,\,\, \nabla^\mu A_\mu=0$ by using the matter field equations.

\section{CCJ in Curved Space}
To include the improvement term in curved space a modification of the minimal coupling to gravity is required. This coupling is called a ``conformally coupled" scalar field
\begin{equation}
(\sqrt{-g})^{-1}{\cal L}={\cal L}_M- \tfrac{1}{12}\varphi^{i\,2} R +{\cal L}_G\label{LMR}\, ,
\end{equation}
where
\begin{eqnarray}
 {\cal L}_M=\ft 12 \partial_\mu \varphi^i\partial_\nu \varphi^i g^{\mu\nu}-V(\varphi),\quad {\cal L}_G=\ft 12  \kappa^{-2}R\, ,
\end{eqnarray} 
and $\kappa^{-1}=m_P=2.4\times 10^{18}$ GeV  is the Planck mass. The Einstein equation derived from the above Lagrangians are
\begin{equation}
\frac{\kappa^{-2}}{2}G_{\mu\nu}=-\frac{1}{2}\Theta_{\mu\nu}+\frac{1}{12}G_{\mu\nu}\varphi^{i\,2}=-\frac{1}{2}\Theta_{\mu\nu}^c\, ,
\end{equation}
with
\begin{equation}
\Theta_{\mu\nu}^c=T_{\mu\nu}-\frac{1}{6}\left(\nabla_\mu\partial_\nu-g_{\mu\nu}\nabla^2\right)\varphi^{i\,2}-\frac{1}{6}G_{\mu\nu}\varphi^{i\,2}\, .
\end{equation}
$\Theta_{\mu\nu}^c$ is covariantly conserved: $\nabla^\mu \Theta_{\mu\nu}^c=0$ as a consequence of the modified matter field  equations. 
Note that the term $-\ft13 G_{\mu\nu}\partial^\mu\varphi^i \varphi^i$ in $\nabla^\mu \Theta_{\mu\nu}^c$ is cancelled by a 
term coming from 
$\nabla^\mu T_{\mu\nu}$, which cancels the $\ft 16 R \partial_\nu \varphi^i \varphi^i$ and a 
term coming from $-\ft16 (\nabla^2 \partial_\nu -\partial_\nu \nabla^2)\varphi^{i\,2}$, which cancels the
$-\ft13 R_{\mu\nu}\partial^\mu \varphi^i \varphi^i$ term. If the matter system is conformal , then 
\begin{equation}
\Theta_\mu^{c\,\,\mu}=0 \, ,
\end{equation}
which implies $R=0$.

\section{CCJ and Supergravity}
A further modification of $\Theta_{\mu\nu}^c$ occurs if the matter fields $\varphi^i$ are coupled  to a $U(1)$ gauge field, as it occurs in supergravity theory due to the $U(1)$ R-symmetry of the $N=1$ superconformal algebra. In this case the first term in $\Theta_{\mu\nu}^c$, i.e. the canonical stress tensor $T_{\mu\nu}$ contains the gauge fields, which, through  the equations of motion of the scalar field, give an extra term in the conservation equation $\nabla^\mu\Theta_{\mu\nu}^c=J^\mu(\varphi, A) F_{\mu\nu}(A)$. Thanks to its equations of motion, the auxiliary field 
$A_\mu=J_\mu(\varphi, A)$ produces a term which is exactly cancelled by the two terms obtained by varying the supergravity action
\begin{equation}
\nabla^\mu\left(\frac{2}{\sqrt{-g}}\frac{\delta {\cal L}_{SG}}{\delta g^{\mu\nu}}\right)=6 A^\mu F_{\mu\nu}(A)\, .
\end{equation}
In other words if we add to $\Theta_{\mu\nu}^c$ a term proportional to $A_\mu A_\nu -\ft 12 g_{\mu\nu} A_\rho^2$ then $\nabla^\mu \hat{\Theta}_{\mu\nu}^c=0$, which is consistent with the generalized Einstein equations
\begin{align}
&\frac{1}{2\kappa^2}G_{\mu\nu}=-\frac{1}{2}\hat{\Theta}_{\mu\nu}^c\, ,\\
&\hat{\Theta}_{\mu\nu}^c=T_{\mu\nu}-\ft16\left(\nabla_\mu\partial_\nu-g_{\mu\nu}\nabla^\rho \nabla_\rho \right)\varphi^i \bar \varphi^i-\ft16G_{\mu\nu}\varphi^i \bar \varphi^i+A_\mu A_\nu-\ft12g_{\mu\nu} A_\rho^2\, .
\end{align}

\section{Noether Currents for the Superconformal Algebra}

Conformal symmetry in flat space implies Weyl symmetry in curved space. For the vierbein $e_{a\mu}(x)$, the electromagnetic field $F_{\mu\nu}$, and the scalar field $\phi$, the Weyl transformations are
\begin{align}
&e_{a\mu}^\prime=e^{\lambda(x)}e_{a\mu}\, ,\\
&F_{\mu\nu}^\prime=F_{\mu\nu}\, ,\\
&\phi^{i\,\prime}(x)=e^{-\lambda (x)}\phi^i(x)\, ,\\
&R^\prime=e^{-2\lambda(x)}R +\partial \lambda \,\, \mbox{terms}\, .
\end{align}
So for zero scalar masses the electromagnetic system and the scalars (coupled to gauge fields) are Weyl invariant. 

The Weyl symmetry of a scalar system actually allows to consider one of them as non-dynamical, since if $\phi(x)\neq 0$ we can choose the Weyl parameter such that 
$\phi(x)=e^{\lambda(x)}$ and set $\phi^\prime(x)=1$.  Then this procedure introduces a scale $\kappa^{-1}=m_P$. In the absence of the Einstein term  ${\cal L}_G$, the scalar Lagrangian \eqref{LMR} becomes ${\cal L}(\phi^\prime=\kappa^{-1})=-1/12\kappa^{-2}R$ using Weyl symmetry (since ${\cal L}_M(\phi^\prime)=0$ at $\partial \phi^\prime=0$). 
This gives the Einstein term with a wrong sign. So, we must introduce a matter field $\phi_0$, with the wrong sign of the kinetic term, ${\cal L}_{\phi_0}=-\ft 12 \partial_\mu \phi_0\partial_\nu \phi_0 g^{\mu\nu}+\frac{1}{12}\phi_0^2 R-\lambda \phi_0^4$. Notice that the self-interaction term $\lambda \phi_0^4$ introduces a cosmological constant after gauge fixing.

In analogy with the Einstein action obtained by a gauge fixed Weyl action we can obtain supergravity by a chiral multiplet action, which is superconformal invariant in flat space and super Weyl invariant in curved superspace, after gauge fixing $X^0=\left(\kappa^{-1}, \psi^0=0, F^0=u\right)$\footnote{Note that the $u$ of \cite{Ferrara:2017yhz} is identified with $\bar u$ of \cite{Cremmer:1982en}.}. 

The superconformal algebra in flat space is a simple superalgebra (in mathematical terms) $SU(2,2/1)$ with (Lie Algebra) part $SU(2,2)\times U_R(1)$. Here $SU(2,2)\sim SO(4,2)$ and the R-symmetry is part of the algebra. The odd supersymmetry generators are $Q_\alpha, S_\alpha$. Their anticommutators generate the even-part of the superalgebra
\begin{align}
&\{Q,\bar Q \} \to P_\mu\, , \qquad \qquad \quad \, \{S,\bar S\} \to K_\mu\, ,\\
&\{Q, \bar S \}=\{\bar Q,S\}=0\, , \qquad \{Q,S\} \to M_{\mu\nu}, D, \Pi \, ,\\
&\{Q,D\} \to \ft 12 Q\, , \qquad \qquad \quad \{S,D\}\to -\ft12 S^\prime\, ,\\
&\{Q,\Pi  \} \to Q\, ,  \qquad \qquad \quad  \,\,\,\, \{S, \Pi \} \to -S\, ,\\
&[Q,P]\to 0\, , \qquad \qquad \qquad \quad[S,K]\to 0\, ,\\
&[Q,K_\mu] \to S\, , \qquad \qquad \qquad  \,\, [S,P]\to Q\, .
\end{align}

The Noether currents of the supercharges $(Q,S)$ can be written in a unified way by introducing an $x$-dependent (anticommuting) spinor parameter 
$\epsilon(x)=\epsilon_0+\slashed{x}\epsilon_1$ and  by writing the spinor Noether currents for $Q$ and $S$ supersymmetry as $\bar\epsilon ^\alpha(x) J_{\alpha\mu}$. Its conservation implies
\begin{eqnarray}
\epsilon_0\,\, \mbox{term}\to \partial^\mu J_{\alpha\mu}=0 \quad \epsilon_1 \,\,\mbox{term}\to \partial^\mu(\slashed{x} J_\mu)=0 \to \gamma^\mu J_{\mu\alpha}=0\, .
\end{eqnarray}
The $Q_\alpha, S_\alpha $ generators are evaluated from the expression $\bar \epsilon_0 Q+\bar \epsilon_1 S\to \int d^3 x\bar \epsilon^\alpha(x)J_{\alpha 0}(x)$.
So, two superconformal transformations, of parameters $\epsilon(x), \eta(x)$ generate a space-time conformal transformation of parameter $\xi^\mu(x)=2 i \bar \epsilon (x)\gamma^\mu \eta(x)$, which indeed satisfies $\partial_\mu \xi_\nu+\partial_\nu \xi_\mu=\ft 12 \eta_{\mu\nu}\partial^\lambda \xi_\lambda$ as a consequence of $\epsilon(x)=\epsilon_0+\slashed{x}\epsilon_1, \eta(x)=\eta_0+\slashed{x}\eta_1$ \cite{Wess:1974tw,Wess:1974kz}.

\section{The Supercurrent and Super Conservation Laws}
The Lagrangian with an improved energy-momentum tensor, in curved space reads as \cite{Callan:1970ze}
\begin{equation}
{\cal L}^M_{improved}=\sqrt{-g}\left( \frac{1}{2}\partial_\mu \phi^i\partial_\nu \bar \phi^i-\frac{1}{12}R \phi^i \bar \phi^i \right) -\sqrt{-g} V(\phi)\, .
\end{equation}
Its Einstein equations are
\begin{equation}
\frac{1}{2\kappa^2}\left(R_{\mu\nu}-\ft 12 g_{\mu\nu} R \right)=\frac{1}{2\kappa^2} G_{\mu\nu}=-\ft 12 T_{\mu\nu}^{improved}=-\ft 12 \Theta_{\mu\nu}^c\, ,
\end{equation}
and $T_{\mu\nu}^M=-\tfrac{1}{2\sqrt{-g}}\tfrac{\delta {\cal L}^M}{\delta g^{\mu\nu}}$. In supergravity these equations are modified because of the $R$-symmetry in the superconformal algebra. 

Let us give some examples \cite{Ferrara:1974pz} of supercurrent multiplets which contain the improved energy-mometnum tensor $\Theta_{\mu\nu}$ (in flat space). The supercurrent for a chiral multiplet is
\begin{equation}
J_{\alpha\dot\alpha}^s=i S^i \sigma^\mu_{\alpha\dot\alpha} \overleftrightarrow{\partial_\mu}\bar S^i + \ft 12 D_\alpha S\overline D_{\dot\alpha}\bar S^i \, .
\end{equation}
The supercurrent for a vector multiplet is
\begin{equation}
J^V_{\alpha\dot\alpha}=W_\alpha\overline W_{\dot\alpha}\, , \qquad (W_\alpha=\overline D^2 D_\alpha V) \, .
\end{equation}
They both obey $\overline D^{\dot\alpha} J_{\alpha\dot\alpha}=0$ as a consequence of the field equations. In the Maxwell case this follows from $\overline D^{\dot\alpha}\overline W_{\dot\alpha}=0$. In the non-conformal case the supercurrent for chiral multiplets with canonical kinetic term and superpotential $W$ satisfies
\begin{equation}
\overline D^{\dot\alpha} \sigma^\mu_{\alpha\dot\alpha}J_{\mu}(x, \theta, \bar \theta) =D_\alpha Y \qquad (\overline D_{\dot\alpha} Y=0)\, ,
\end{equation}
 with 
 \begin{equation}
 Y=(X^0)^3(W-\ft13 W_iS^i) =(X^0)^3 \Delta W \, ,
 \end{equation}
where $Y=0$ for cubic $W$. Note that in the local superconformal formulation $W$ must be a function of degree 3 in $X^I=(X^0 S^i, X^0)$, so that ${\cal W}= (X^0)^3W(S^i)$.
The $Y$ chiral multiplet satisfying the (partial) conservation equation is made of fields $(Y,\, \Psi_Y, \,\, F_Y)$ of physical dimensions $(3,7/2,4)$. Its components are 
\begin{eqnarray}
Y=(X^0)^3(W-\ft13W_i S^i), \,\, \Psi_Y=(\gamma^\mu J_\mu)_\alpha, \,\, F_Y=\Theta_\lambda ^{\,\,\lambda}+i \partial^\mu J_\mu^5 \, .
\end{eqnarray}
Note that $Y=0$ in the superconformal case, where $W$ is of degree 3 in $S^i$.
In supergravity the multiplet which contains the stress tensor, the supercurrent and the $R$-symmetry current
\begin{eqnarray}
J_\mu^5(x), \Theta_{\mu\nu}(x), J_{\mu\alpha}(x)\to  J_\mu^5+i \theta^\alpha J_{\mu\alpha}+i \bar \theta_{\dot\alpha} J_\mu^{\dot\alpha}+\theta \sigma^\nu \bar \theta\left( \Theta_{\mu\nu}+\epsilon_{\mu\nu\rho\sigma} \partial^\rho J^{5\, \sigma}\right)+... \, ,
\end{eqnarray}
should couple to the supergravity fields.
When $Y\neq 0$ superconformal symmetry is broken.  In supergravity, non-canonical kinetic terms may also contribute to $Y$ \cite{Ferrara:2017yhz}. For  some particular choices $Y=0$ as is the case for conformally coupled scalars \cite{Callan:1970ze}.

\section{Superspace Description of Supergravity}
The basic fields of the supergravity \cite{Ferrara:1978em,Stelle:1978ye,Ferrara:1978wj,Stelle:1978yr} are the vierbein, the gravitino and the auxiliary fields
\begin{align}
&e_{a\mu} \quad (\mbox{vierbein field})\, , \,\, g_{\mu\nu}=e_{a\mu}e_{b\nu}\eta^{ab}\, , \\
&\psi_{\mu\alpha} \quad \mbox{gravitino} \, ,\\
&A_\mu,\,\, u: \quad \mbox{auxiliary fields}\, .
\end{align}

In superconformal gravity the gauge fields $e_{a\mu}, \psi_{\mu\alpha}, A_\mu$ gauge the superconformal algebra (super Weyl symmetry in superspace).
$u$ is the complex scalar residual of the gauge fixed superconformal compensator. $X^0$ set to $(\kappa^{-1}, 0, u)$ to gauge fix the super Weyl symmetry. $A_\mu$ is the $R$-symmetry gauge field, whose gauge invariance is broken in Einstein supergravity, so that it has four independent components. 

The geometry of superspace is encoded in three basic multiplets \cite{Ferrara:1977mv,Wess:1978ns,Wess:1978bu,Wess:1977fn,Gates:1983nr,Townsend:1979js,Ferrara:1988qx} which contain the superspace curvatures. These multiplets are denoted as ${\cal R}$, $E_\mu$, $W_{\alpha\beta\gamma}$. The chiral curvature multiplet ${\cal R}$ contains in the last $\theta^2$ component the scalar curvature $R$ through $B_\mu^{\,\,\mu}$, the real Einstein multiplet $E_\mu$ contains the Einstein tensor $G_{\mu\nu}$ through $B_{\mu\nu}$  in the $\theta\bar\theta$ component. The Weyl tensor $C_{\alpha\beta\gamma\delta}$ (traceless part of the Riemann tensor) is contained in the $\theta$ component of the chiral Weyl multiplet $W_{\alpha\beta\gamma} $. The relevant formulae are given below. 
\begin{align}
& {\cal R}=\bar u + \theta^\alpha \gamma^\mu  R_\mu +  \theta^2\left( -\ft 16B_\mu^{\,\,\mu}-i D^\mu A_\mu\right)\, ,\\
&E_\mu=\sigma_\mu^{\alpha\dot\alpha} E_{\alpha\dot\alpha}=A_\mu+\theta^\alpha Z_{\mu\alpha}+\theta^\alpha \sigma_{\alpha\dot\alpha}^\nu B_{\mu\nu}\bar \theta^{\dot\alpha}+...\, ,\\
&W_{\alpha\beta\gamma}=...\theta^\delta (C_{\alpha\beta\gamma\delta}+F_{(\alpha\beta}\epsilon_{\gamma)\delta})+\theta^2 (\mbox{fermion})\, ,
\end{align}
where
\begin{equation}
B_{\mu\nu}=3R_{\mu\nu}-\ft12g_{\mu\nu}R-6A_\mu A_\nu +3 g_{\mu\nu} A_\rho^ 2+3g_{\mu\nu}u \bar u\, .
\end{equation}
The $x_\mu$, $\theta_\alpha$ are superspace coordinates satisfying
\begin{eqnarray}
x_\mu x_\nu=x_\nu x_\mu,\quad x_\mu \theta_\alpha =\theta_\alpha x_\mu, \quad \theta_\alpha \theta_\beta =- \theta_\beta\theta_\alpha    \, .
\end{eqnarray}

\section{Superconformal Matter and Supercurrents in Curved Superspace}
The Einstein equations come from a supermultiplet of equations 
\begin{equation}
{\cal E}_{\alpha\dot\alpha}+ J_{\alpha\dot\alpha}=0\, , \label{EJ} 
\end{equation}
whose first component is $\frac{\delta {\cal L}}{\delta A_{\alpha\dot\alpha}}=0$. The trace and conservation of the Einstein tensor come from the fundamental superspace (off-shell) identity
\begin{equation}
\overline D^{\dot\alpha} {\cal E}_{\alpha\dot\alpha}= (X^0)^3 D_\alpha \left( \frac{{\cal R}}{X^0} \right)\, , \label{calER}
\end{equation}
and (on-shell) supercurrent conservation
\begin{equation}
\overline D^{\dot\alpha} J_{\alpha\dot\alpha}= (X^0)^3 D_\alpha \left( \frac{Y}{(X^0)^3} \right)\, , 
\end{equation}
which, {\it{up to a constant}}, imply
\begin{equation}
{\cal R}+\frac{Y}{(X^0)^2}=0 \, . 
\end{equation}
In the superconformal gauge $X^0=\kappa^{-1}$ these equations read \footnote{Note that ${\cal E}_{\alpha\dot\alpha}$ has conformal weight 3, while $E_{\alpha\dot\alpha}$ and ${\cal R}$ have physical dimension 1. Therefore, the physical dimension of all  auxiliary fields is one.}
\begin{eqnarray}
E_{\alpha\dot\alpha}=-\kappa^2 J_{\alpha\dot\alpha}, \qquad \overline D^{\dot\alpha}E_{\alpha\dot\alpha}=D_\alpha {\cal R}, \qquad \overline{D}^{\dot\alpha} J_{\alpha\dot\alpha}=D_\alpha Y, \qquad {\cal R}+ \kappa^2 Y=0\, . 
\end{eqnarray}
They contain the GR identity
 \begin{equation}
 \nabla^\mu G_{\mu\nu}=0 \, .
 \end{equation}
The first equation is related to the global formulae in \cite{Clark:1995bg,Komargodski:2009rz, Komargodski:2010rb,Kuzenko:2010am,Kuzenko:2010ni,Korovin:2016tsq}
and for superconformal matter yields \footnote{Note that in the superconformal formulation all dimensionfull parameters get their physical dimension from $\kappa$ factors coming from $X^0=\kappa^{-1}$. The remaining dimensionless parameters correspond to conformal couplings. }
\begin{equation}
{\cal R}=Y=0 \, . 
\end{equation}
The constant term is related to a term in $Y$ of the form $\lambda (X^0)^3$ so that ${\cal R}+\lambda X^0=0$. This corresponds to pure anti de Sitter supergravity  \cite{Townsend:1977qa,Ferrara:1978rk}.

In this setup the supergravity Lagrangian is given by the following expression \cite{Ferrara:1978em,Stelle:1978ye}
\begin{equation}
{\cal L}_{SG}(e, \psi, A, u)=\kappa^{-2}\left(\ft 12R-\ft 12 \bar \psi^\mu R_\mu -3 u \bar u+ 3 A_\mu^2\right)\, , 
\end{equation}
so that the bosonic contribution to the (super)Einstein tensor is
\begin{equation}
\frac{1}{\sqrt{-g}}\frac{\delta {\cal L}}{\delta g^{\mu\nu}}=\ft 12 G_{\mu\nu}+\ft 32 g_{\mu\nu} u\bar u +3 A_\mu A_\nu- \ft 32 g_{\mu\nu} A_\rho^2\, . 
\end{equation}
This means that the matter part will have improvement terms to balance the $G_{\mu\nu}$ term in $\nabla^\mu G_{\mu\nu}=0$,  $\nabla^\mu(G_{\mu\nu}+...)=0$.

In pure supergravity $A_\mu=u=0$, but in matter coupled systems the improved stress tensor will get supergravity corrections
\begin{equation}
\hat \Theta_{\mu\nu}^c=\Theta_{\mu\nu}^c +\left ( 6 A_\mu A_\nu -3 g_{\mu\nu} A_\rho^2 +3 g_{\mu\nu} u\bar u\right ) \kappa^{-2}\, .\\
\end{equation}
These were found in \cite{Ferrara:2017yhz}. They are such that 
\begin{equation}
\kappa^{-2}G_{\mu\nu}+\hat\Theta_{\mu\nu}^c=0\, ,
\end{equation}
will be consistent with the matter conservation laws
\begin{equation}
\nabla^\mu \hat \Theta_{\mu\nu}^c=0\, ,
\end{equation}
with the additional property $\hat{\Theta}_\mu^{c\,\,\mu}=0$ in the superconformal case.

\section{Rigid Supersymmetry Breaking}
Given a generic ``superfield" $\phi(x, \theta)$ with  a given $n_{last}$, its component expression is (symbolically)
\begin{align}
\phi(x,\theta)=\phi_0(x)+\theta \phi_1(x)+...
+\theta^{n-1}\phi_{n-1}(x)+\theta^n\phi_n(x)+\theta^{n+1}\phi_{n+1}(x)+... \theta^{n_{last}}\phi_{last}(x)\, .
\end{align}
The supersymmetry transformation of the $\theta^n$ component is
\begin{align}
\delta_\epsilon \phi_n(x)=\epsilon \partial_x \phi_{n-1}(x)+\epsilon \phi_{n+1}(x)\, .
\end{align}
So if $ <\phi_{n+1}>\neq 0$ then supersymmetry is broken
\begin{eqnarray}
<\delta_\epsilon \phi_n(x)>=<\epsilon\{ Q, \phi_n(x)]>\neq 0 \quad \Rightarrow \quad Q|0>\neq 0\, . 
\end{eqnarray}
Thus the only field which can have a supersymmetry preserving vacuum expectation value is $\phi_0(x)$.
This argument, applied to the supercurrent multiplet, implies that supersymmetry is broken if $\left<\Theta_{\mu\nu}\right>\neq 0$. Indeed, in a Lorentz and translational invariant vacuum, 
$\left<\Theta_{\mu\nu}\right>=-\eta_{\mu\nu}\left<V\right>\geq 0$, where $V$ is the scalar potential. Hence\footnote{$V_0=V|_{\partial V/\partial\phi =0}$ is an extremum of the potential.}, $V_0>0$ corresponds to broken supersymmetry.
A supersymmetry preserving vacuum has $V_0=0$. 

This argument is not valid in curved space as we know that $AdS$ (Anti de Sitter) can be a supersymmetry preserving vacuum.

\section{Rigid Curved Supersymmetry, SuperHiggs Effect and the Cosmological Constant}

Rigid vacua in curved geometries have recently been studied in many papers starting with \cite{Festuccia:2011ws,Dumitrescu:2012ha}. For vacua preserving maximal (4 charges) supersymmetry in curved $4d$ space, an efficient method to find them is to look for the solution of the curved superspace identity
\begin{equation}
\overline D^{\dot\alpha}E_{\alpha\dot\alpha}= D_\alpha {\cal R}\, ,
\end{equation} 
and ask for vacua for which higher $\theta$ components of the geometric superfields ${\cal R},\, E_{\alpha\dot\alpha}, \, W_{\alpha\beta\gamma}$ are vanishing, but the lowest components are not. 
Other than Minkowski space ($u=A_\mu=0$) one finds two vacua with four supersymmetries: $AdS_4=\tfrac{SO(3,2)}{SO(3,1)}$ and  $S^3 \times L$. For the first case 
\begin{equation}
{\cal R}=\left(\left<\bar u\right>\neq 0, 0, 0\right),
\end{equation}
which hence satisfies
\begin{equation}
E_{\alpha\dot\alpha} =0 \quad W_{\alpha\beta\gamma}=0\, .
\end{equation}
In this case the Ricci tensor is $R_{\mu\nu}=-3 g_{\mu\nu}|u|^2$ 
and it is obtained by $B_\mu^{\,\,\mu}=0$.
The $S^3 \times L$ manifold is given by
\begin{equation}
 \sigma_\mu^{\alpha\dot\alpha} E_{\alpha\dot\alpha} =\left(\left< A_\mu\right>, 0, 0,...,0 \right) \, ,
\end{equation}
where one should choose $\left<A_\mu\right>=(A_0, \overrightarrow{0})$.
Both solutions satisfy $\overline D^{\dot\alpha} E_{\alpha\dot\alpha}=D_\alpha {\cal R}=0$.
The Einstein curvature of these two spaces is retrieved by the vanishing of upper components of ${\cal R}$ and $E_{\alpha\dot\alpha}$, respectively given by
\begin{equation}
\ft 16  R+A_\mu^2+ 2 u \bar u =\ft 16 B_\mu^{\,\,\mu} \quad \mbox{and} \quad B_{\mu\nu}\, .
\end{equation}
For $S^3\times L$,  $u=0$ and the Ricci tensor $R_{\mu\nu}$ is computed by the equation $B_{\mu\nu}=0$. For the non trace part we have
\begin{equation}
R_{00}=R_{0i}=0, \qquad R_{ij}=2 \delta_{i j}A_0^2 \, . 
\end{equation}
These spaces also satisfy \cite{Festuccia:2011ws,Dumitrescu:2012ha,Ferrara:2017yhz}
\begin{align}
%& B_{\mu\nu}=0 \qquad \mbox{vanishing of upper components of} \,\, E_{\alpha\dot\alpha}\\
& u A_\mu=0, \quad \, \nabla_\mu A_\nu=0, \quad \partial_\mu u =0\, ,\label{uAmu}\\
&C_{\alpha\beta\gamma\delta}=0  \quad \mbox{(conformally flat spaces)} \label{Cabg}\, .
\end{align}
The equations \eqref{uAmu} and  \eqref{Cabg} define supersymmetric curved backgrounds then satisfying $\overline D^{\dot\alpha} {\cal E}_{\alpha\dot\alpha}= (X^0)^3 D_\alpha \left( {\cal R}/{X^0} \right)=0$  and $W_{\alpha\beta\gamma}=0$  \cite{Ferrara:2017yhz}. 
In a  similar fashion one can prove that the other maximally symmetric space $dS=\tfrac{SO(4,1)}{SO(3,1)}$ is not supersymmetric \cite{Antoniadis:2014oya,Bergshoeff:2015tra,Hasegawa:2015bza}. Indeed for this space the upper components of the chiral scalar curvature multiplet ${\cal R}=...\theta^2\left(-\ft 16 R-2 u\bar u\right)$ is not vanishing.
If one combines the ansatz for the $AdS$ and $dS$ curvatures, which give a cosmological constant, 
\begin{equation}
V(\mu,\lambda)=\ft 13 \kappa^{-4}\left(\mu^2-9 \lambda^2\right)\, ,
\end{equation}
one gets for the first and last component of the scalar curvature multiplet ${\cal R}$, respectively.
\begin{equation}
\kappa{\cal R}|_{first}= \lambda=\kappa^2 F^0,  \qquad \kappa {\cal R}|_{last}= -\ft 29 \kappa^{-1}\mu^2 = -2 \kappa^3 (F^1)^2\, .
\end{equation}
These equations show that the $\mu$ term breaks supersymmetry. $F^0=\kappa^{-1} u$ and $F^1=-\ft 13 \kappa^{-2}\mu$ is the auxiliary field of the goldstino multiplet. Depending on whether $\mu^2\le 9\lambda^2,\,\, \mu^2> 9\lambda^2$ these configurations break SUSY (SuperHiggs effect) in $AdS$, Minkowski and $dS$. SUSY is unbroken whenever $\mu=0$. 

\section{No-scale Supergravity}
In the conformal setting no-scale supergravity \cite{Cremmer:1983bf} arises as a particular deformation $\Delta W$ of the conformal (cubic) superpotential
\begin{eqnarray}
&& W=\ft 12 (\sigma +S)^3\label{Wsi}\, ,\\
&&\sigma W_S= 3\Delta W =3 W- S W_S\, .\label{WsiS}
\end{eqnarray}
Using the general formula for the potential
\begin{equation}
V=\frac{\kappa^{-4}}{3 }\left( (\Phi_M^{-1})^{i \bar j} W_i \overline W_{\bar j} -|3\Delta W|^2 \right)
\end{equation}
and plugging in \eqref{WsiS} and \eqref{Wsi} one finds for the scalar potential
\begin{equation}
V=\frac{\kappa^{-4}}{3 }\ W_S \overline W_{\bar S}(1-|\sigma |^2)\, .
\end{equation}
For $\sigma =1, \,\, V=0$ with $W_S\neq 0$ and the no-scale structure is obtained \cite{Cremmer:1983bf}.

\section{Summary and Conclusions}

We obtain the Einstein equations for matter-coupled supergravity in the conformal tensor calculus formalism. We paid special attention to what we called the conformal case. This is the supergravity coupling of ${\cal N}=1$ rigid supersymmetric models of chiral multiplets with conformal symmetry. In this case the K\"{a}hler couplings imply that there is a $\U(1)$ isometry group (the $R$-symmetry).\footnote{In the simplest case of the conformally coupled scalar there is an additional $\SU(N)$ symmetry, which is not present in the other models satisfying the conformal restriction.}

In \cite{Ferrara:2017yhz} it has been relevant  to consider the difference between two gauge choices for super Weyl symmetry.
In the Einstein gauge  the scalar fields parametrize a K\"{a}hler $\sigma$-model with K\"{a}hler potential $K(S,\bar S)=-3 \log(-\Phi(S,\bar S)/3)$. The conformal case is  characterized by a function $\Phi(S,\bar S)+3$ of degree 1 both in $S$ and $\bar S$, and of the superpotential $W(S)$ of degree $3$.

A conformal gauge preserves the separation between the pure supergravity part, where the superconformal symmetry is broken in order to get super-Poincar\'{e} gravity, and the matter part with preserved conformal symmetry. This separation is maintained by not eliminating the auxiliary gauge field $A_\mu $ of the  $\U(1)$ $R$-symmetry. Then the matter part has still K\"{a}hler couplings, where now the K\"{a}hler potential is $\Phi(S,\bar S)$. These results provide a supersymmetric generalization of the properties of scalar fields coupled to gravity with improvement terms in CCJ \cite{Callan:1970ze}.
Two kind of bosonic improvement terms emerge, one that couples the scalar fields to the scalar curvature $R$, the other that couples the scalar fields to an $R$-current.
Both are part of the superconformal covariant derivatives that covariantize the (rigid conformal) CCJ improvement terms.
Therefore, the improved energy-momentum tensor that is traceless for superconformal matter contains also $\U(1)$ corrections. This also implies an improvement term in the $\U(1)$ current. These are part of the supercurrent, which becomes $\gamma$-traceless in the superconformal case \cite{Ferrara:1974pz}  for which the compensator equation becomes the chiral superfield equation ${\cal R}\approx 0$. We clarify the bosonic aspects, which provides the improved currents for conformal K\"{a}hler couplings \cite{Ferrara:2017yhz}.

We have given explicit formulae, in the superconformal approach, for the three basic multiplets that specify the superspace geometry of ${\cal N}=1$ supersymmetry. These multiplets play a key role in the construction of higher curvature invariants and they have found applications to classify counterterms \cite{Stelle:2012zz,Bern:2014dlu,Deser:1977nt,Kallosh:1980fi}. More recently they were also relevant in cosmology to provide a generalization of the Starobinsky model \cite{Cecotti:1987sa,Cecotti:1987qe,Kallosh:2013xya,Ellis:2013xoa,Farakos:2013cqa,Ferrara:2013rsa} as well as for nonlinear realizations for local supersymmetry in the framework of ${\cal N}=1$ supergravity \cite{Antoniadis:2014oya,Ferrara:2014kva,Ferrara:2015cwa,Bergshoeff:2016psz,Ferrara:2016ajl}. The latter is a particular way for implementing the super-Brout-Englert-Higgs effect and to find de Sitter vacua in cosmological scenarios. It is likely that our results will find new applications along this area of research.

Our results can also be relevant in exploring the interplay between different supergravity backgrounds, in the study of rigid supersymmetry in curved space. The simplest examples, preserving four supersymmetries were discussed in \cite{Ferrara:2017yhz} and correspond to the conformally flat spaces $AdS_4$ and $S^3\times L$. Similar arguments show that the $dS$ background is not supersymmetric.  Another related topic is the application of localization techniques in supersymmetric quantum field theories \cite{Festuccia:2011ws, Dumitrescu:2012ha, Pestun:2016zxk, Pufu:2016zxm}.

\section*{Acknowledgments}
We thank M. Porrati and A. Sagnotti for very useful comments on the manuscript. We especially thank A. van Proeyen and M. Tournoy for a collaboration on the topics of this review.  
The work of S.F. is supported in part by CERN TH Department and INFN-CSN4-GSS.

\providecommand{\href}[2]{#2}\begingroup\raggedright\endgroup


\begin{thebibliography}{10}

\bibitem{Callan:1970ze}
C.~G. Callan, Jr., S.~R. Coleman  and R.~Jackiw, \emph{{A new improved energy -
  momentum tensor}}, Annals Phys. {\bf 59} (1970)
\href{http://dx.doi.org/10.1016/0003-4916(70)90394-5}{42--73}
%%CITATION = APNYA,59,42;%%.

\bibitem{Ferrara:2017yhz}
S.~Ferrara, M.~Samsonyan, M.~Tournoy  and A.~Van~Proeyen, \emph{{The
  supercurrent and Einstein equations in the superconformal formulation}},
\href{http://arxiv.org/abs/1705.02272}{{\tt arXiv:1705.02272 [hep-th]}}
%%CITATION = ARXIV:1705.02272;%%.

\bibitem{Ferrara:1974pz}
S.~Ferrara and B.~Zumino, \emph{{Transformation Properties of the
  Supercurrent}}, Nucl. Phys. {\bf B87} (1975)
\href{http://dx.doi.org/10.1016/0550-3213(75)90063-2}{207}
%%CITATION = NUPHA,B87,207;%%.

\bibitem{Ferrara:1978jt}
S.~Ferrara and P.~van Nieuwenhuizen, \emph{Tensor calculus for supergravity},
  Phys. Lett. {\bf B76} (1978)
\href{http://dx.doi.org/10.1016/0370-2693(78)90893-6}{404}
%%CITATION = PHLTA,76B,404;%%.

\bibitem{Cremmer:1982en}
E.~Cremmer, S.~Ferrara, L.~Girardello  and A.~Van~Proeyen, \emph{{Yang--Mills
  theories with local supersymmetry: Lagrangian, transformation laws and
  superhiggs effect}}, Nucl. Phys. {\bf B212} (1983)
\href{http://dx.doi.org/10.1016/0550-3213(83)90679-X}{413}
%%CITATION = NUPHA,B212,413;%%.

\bibitem{Freedman:2012zz}
D.~Z. Freedman and A.~Van~Proeyen, {\em Supergravity}.
\newblock Cambridge University Press,
2012.
\newblock
%%CITATION = INSPIRE-1123253;%%.

\bibitem{Ferrara:1978em}
S.~Ferrara and P.~van Nieuwenhuizen, \emph{{The auxiliary fields of
  supergravity}}, Phys. Lett. {\bf B74} (1978)
\href{http://dx.doi.org/10.1016/0370-2693(78)90670-6}{333}
%%CITATION = PHLTA,B74,333;%%.

\bibitem{Stelle:1978ye}
K.~S. Stelle and P.~C. West, \emph{{Minimal auxiliary fields for
  supergravity}}, Phys. Lett. {\bf B74} (1978)
\href{http://dx.doi.org/10.1016/0370-2693(78)90669-X}{330}
%%CITATION = PHLTA,B74,330;%%.

\bibitem{Wess:1974tw}
J.~Wess and B.~Zumino, \emph{{Supergauge transformations in four-dimensions}},
  Nucl. Phys. {\bf B70} (1974)
39--50
%%CITATION = NUPHA,B70,39;%%.

\bibitem{Wess:1974kz}
J.~Wess and B.~Zumino, \emph{A Lagrangian model invariant under supergauge
  transformations}, Phys. Lett. {\bf B49} (1974)
52
%%CITATION = PHLTA,B49,52;%%.

\bibitem{Ferrara:1978wj}
S.~Ferrara and P.~Van~Nieuwenhuizen, \emph{{Structure of supergravity}}, Phys.
  Lett. {\bf B78} (1978)
\href{http://dx.doi.org/10.1016/0370-2693(78)90642-1}{573}
%%CITATION = PHLTA,B78,573;%%.

\bibitem{Stelle:1978yr}
K.~S. Stelle and P.~C. West, \emph{{Tensor calculus for the vector multiplet
  coupled to supergravity}}, Phys. Lett. {\bf B77} (1978)
\href{http://dx.doi.org/10.1016/0370-2693(78)90581-6}{376}
%%CITATION = PHLTA,B77,376;%%.

\bibitem{Ferrara:1977mv}
S.~Ferrara and B.~Zumino, \emph{{Structure of linearized supergravity and
  conformal supergravity}}, Nucl. Phys. {\bf B134} (1978)
\href{http://dx.doi.org/10.1016/0550-3213(78)90548-5}{301--326}
%%CITATION = NUPHA,B134,301;%%.

\bibitem{Wess:1978ns}
J.~Wess and B.~Zumino, \emph{{The component formalism follows from the
  superspace formulation of supergravity}}, Phys. Lett. {\bf 79B} (1978)
\href{http://dx.doi.org/10.1016/0370-2693(78)90390-8}{394--398}
%%CITATION = PHLTA,79B,394;%%.

\bibitem{Wess:1978bu}
J.~Wess and B.~Zumino, \emph{{Superfield lagrangian for supergravity}}, Phys.
  Lett. {\bf B74} (1978)
\href{http://dx.doi.org/10.1016/0370-2693(78)90057-6}{51}
%%CITATION = PHLTA,B74,51;%%.

\bibitem{Wess:1977fn}
J.~Wess and B.~Zumino, \emph{{Superspace formulation of supergravity}}, Phys.
  Lett. {\bf B66} (1977)
\href{http://dx.doi.org/10.1016/0370-2693(77)90015-6}{361--364}
%%CITATION = PHLTA,B66,361;%%.

\bibitem{Gates:1983nr}
S.~J. Gates~Jr., M.~T. Grisaru, M.~Ro\v{c}ek  and W.~Siegel, \emph{{Superspace,
  or one thousand and one lessons in supersymmetry}}, Front. Phys. {\bf 58}
  (1983) 1--548,
\href{http://arxiv.org/abs/hep-th/0108200}{{\tt arXiv:hep-th/0108200}}
%%CITATION = HEP-TH/0108200;%%.

\bibitem{Townsend:1979js}
P.~K. Townsend and P.~van Nieuwenhuizen, \emph{{Anomalies, topological
  invariants and the {Gauss--Bonnet} theorem in supergravity}}, Phys. Rev. {\bf
  D19} (1979)
\href{http://dx.doi.org/10.1103/PhysRevD.19.3592}{3592}
%%CITATION = PHRVA,D19,3592;%%.

\bibitem{Ferrara:1988qx}
S.~Ferrara and M.~Villasante, \emph{{Curvatures, Gauss--Bonnet and
  Chern--Simons multiplets in old minimal $N=1$ supergravity}}, J. Math. Phys.
  {\bf 30} (1989)
\href{http://dx.doi.org/10.1063/1.528576}{104}
%%CITATION = JMAPA,30,104;%%.

\bibitem{Clark:1995bg}
T.~E. Clark and S.~T. Love, \emph{{The Supercurrent in supersymmetric field
  theories}}, Int. J. Mod. Phys. {\bf A11} (1996)
  \href{http://dx.doi.org/10.1142/S0217751X9600136X}{2807--2823},
\href{http://arxiv.org/abs/hep-th/9506145}{{\tt arXiv:hep-th/9506145 [hep-th]}}
%%CITATION = HEP-TH/9506145;%%.

\bibitem{Komargodski:2009rz}
Z.~Komargodski and N.~Seiberg, \emph{{From linear SUSY to constrained
  superfields}}, JHEP {\bf 0909} (2009)
  \href{http://dx.doi.org/10.1088/1126-6708/2009/09/066}{066},
\href{http://arxiv.org/abs/0907.2441}{{\tt arXiv:0907.2441 [hep-th]}}
%%CITATION = ARXIV:0907.2441;%%.

\bibitem{Komargodski:2010rb}
Z.~Komargodski and N.~Seiberg, \emph{{Comments on supercurrent multiplets,
  supersymmetric field theories and supergravity}}, JHEP {\bf 1007} (2010)
  \href{http://dx.doi.org/10.1007/JHEP07(2010)017}{017},
\href{http://arxiv.org/abs/1002.2228}{{\tt arXiv:1002.2228 [hep-th]}}
%%CITATION = ARXIV:1002.2228;%%.

\bibitem{Kuzenko:2010am}
S.~M. Kuzenko, \emph{{Variant supercurrent multiplets}}, JHEP {\bf 04} (2010)
  \href{http://dx.doi.org/10.1007/JHEP04(2010)022}{022},
\href{http://arxiv.org/abs/1002.4932}{{\tt arXiv:1002.4932 [hep-th]}}
%%CITATION = ARXIV:1002.4932;%%.

\bibitem{Kuzenko:2010ni}
S.~M. Kuzenko, \emph{{Variant supercurrents and Noether procedure}}, Eur. Phys.
  J. {\bf C71} (2011)
  \href{http://dx.doi.org/10.1140/epjc/s10052-010-1513-1}{1513},
\href{http://arxiv.org/abs/1008.1877}{{\tt arXiv:1008.1877 [hep-th]}}
%%CITATION = ARXIV:1008.1877;%%.

\bibitem{Korovin:2016tsq}
Y.~Korovin, S.~M. Kuzenko  and S.~Theisen, \emph{{The conformal supercurrents
  in diverse dimensions and conserved superconformal currents}}, JHEP {\bf 05}
  (2016) \href{http://dx.doi.org/10.1007/JHEP05(2016)134}{134},
\href{http://arxiv.org/abs/1604.00488}{{\tt arXiv:1604.00488 [hep-th]}}
%%CITATION = ARXIV:1604.00488;%%.

\bibitem{Townsend:1977qa}
P.~K. Townsend, \emph{{Cosmological constant in supergravity}}, Phys. Rev. {\bf
  D15} (1977)
\href{http://dx.doi.org/10.1103/PhysRevD.15.2802}{2802--2804}
%%CITATION = PHRVA,D15,2802;%%.

\bibitem{Ferrara:1978rk}
S.~Ferrara, M.~T. Grisaru  and P.~van Nieuwenhuizen, \emph{{Poincar{\'e} and
  conformal supergravity models with closed algebras}}, Nucl.Phys. {\bf B138}
  (1978)
\href{http://dx.doi.org/10.1016/0550-3213(78)90389-9}{430--444}
%%CITATION = NUPHA,B138,430;%%.

\bibitem{Festuccia:2011ws}
G.~Festuccia and N.~Seiberg, \emph{{Rigid supersymmetric theories in curved
  superspace}}, JHEP {\bf 1106} (2011)
  \href{http://dx.doi.org/10.1007/JHEP06(2011)114}{114},
\href{http://arxiv.org/abs/1105.0689}{{\tt arXiv:1105.0689 [hep-th]}}
%%CITATION = ARXIV:1105.0689;%%.

\bibitem{Dumitrescu:2012ha}
T.~T. Dumitrescu, G.~Festuccia  and N.~Seiberg, \emph{{Exploring curved
  superspace}}, JHEP {\bf 08} (2012)
  \href{http://dx.doi.org/10.1007/JHEP08(2012)141}{141},
\href{http://arxiv.org/abs/1205.1115}{{\tt arXiv:1205.1115 [hep-th]}}
%%CITATION = ARXIV:1205.1115;%%.

\bibitem{Antoniadis:2014oya}
I.~Antoniadis, E.~Dudas, S.~Ferrara  and A.~Sagnotti, \emph{{The
  Volkov-Akulov-Starobinsky supergravity}}, Phys.Lett. {\bf B733} (2014)
  \href{http://dx.doi.org/10.1016/j.physletb.2014.04.015}{32--35},
\href{http://arxiv.org/abs/1403.3269}{{\tt arXiv:1403.3269 [hep-th]}}
%%CITATION = ARXIV:1403.3269;%%.

\bibitem{Bergshoeff:2015tra}
E.~A. Bergshoeff, D.~Z. Freedman, R.~Kallosh  and A.~Van~Proeyen, \emph{{Pure
  de Sitter supergravity}}, Phys. Rev. {\bf D92} (2015), no.~8,
  \href{http://dx.doi.org/10.1103/PhysRevD.93.069901,
  10.1103/PhysRevD.92.085040}{085040},
  \href{http://arxiv.org/abs/1507.08264}{{\tt arXiv:1507.08264 [hep-th]}},
[Erratum: Phys. Rev.D93,no.6,069901(2016)]
%%CITATION = ARXIV:1507.08264;%%.

\bibitem{Hasegawa:2015bza}
F.~Hasegawa and Y.~Yamada, \emph{{Component action of nilpotent multiplet
  coupled to matter in 4 dimensional $ \mathcal{N}=1 $ supergravity}}, JHEP
  {\bf 10} (2015) \href{http://dx.doi.org/10.1007/JHEP10(2015)106}{106},
\href{http://arxiv.org/abs/1507.08619}{{\tt arXiv:1507.08619 [hep-th]}}
%%CITATION = ARXIV:1507.08619;%%.

\bibitem{Cremmer:1983bf}
E.~Cremmer, S.~Ferrara, C.~Kounnas  and D.~V. Nanopoulos, \emph{{Naturally
  vanishing cosmological constant in $N=1$ supergravity}}, Phys. Lett. {\bf
  B133} (1983)
\href{http://dx.doi.org/10.1016/0370-2693(83)90106-5}{61}
% .

\bibitem{Stelle:2012zz}
K.~S. Stelle, \emph{{Ultraviolet infinities and counterterms in supersymmetric
  theories}}, Int. J. Geom. Meth. Mod. Phys. {\bf 09} (2012)
\href{http://dx.doi.org/10.1142/S0219887812610130}{1261013}
%%CITATION = 00436,09,1261013;%%.

\bibitem{Bern:2014dlu}
Z.~Bern, \emph{{Perturbative gravity from gauge theory}}, Mod. Phys. Lett. {\bf
  A29} (2014), no.~32,
\href{http://dx.doi.org/10.1142/S0217732314300365}{1430036}
%%CITATION = MPLAE,A29,1430036;%%.

\bibitem{Deser:1977nt}
S.~Deser, J.~H. Kay  and K.~S. Stelle, \emph{{Renormalizability Properties of
  Supergravity}}, Phys. Rev. Lett. {\bf 38} (1977)
  \href{http://dx.doi.org/10.1103/PhysRevLett.38.527}{527},
\href{http://arxiv.org/abs/1506.03757}{{\tt arXiv:1506.03757 [hep-th]}}
%%CITATION = ARXIV:1506.03757;%%.

\bibitem{Kallosh:1980fi}
R.~E. Kallosh, \emph{{Counterterms in extended supergravities}}, Phys. Lett.
  {\bf 99B} (1981)
\href{http://dx.doi.org/10.1016/0370-2693(81)90964-3}{122--127}
%%CITATION = PHLTA,99B,122;%%.

\bibitem{Cecotti:1987sa}
S.~Cecotti, \emph{{Higher derivative supergravity is equivalent to standard
  supergravity coupled to matter.1.}}, Phys. Lett. {\bf B190} (1987)
\href{http://dx.doi.org/10.1016/0370-2693(87)90844-6}{86--92}
%%CITATION = PHLTA,B190,86;%%.

\bibitem{Cecotti:1987qe}
S.~Cecotti, S.~Ferrara, M.~Porrati  and S.~Sabharwal, \emph{{New minimal higher
  derivative supergravity coupled to matter}}, Nucl. Phys. {\bf B306} (1988)
\href{http://dx.doi.org/10.1016/0550-3213(88)90175-7}{160--180}
%%CITATION = NUPHA,B306,160;%%.

\bibitem{Kallosh:2013xya}
R.~Kallosh and A.~Linde, \emph{{Superconformal generalizations of the
  Starobinsky model}}, JCAP {\bf 1306} (2013)
  \href{http://dx.doi.org/10.1088/1475-7516/2013/06/028}{028},
\href{http://arxiv.org/abs/1306.3214}{{\tt arXiv:1306.3214 [hep-th]}}
%%CITATION = ARXIV:1306.3214;%%.

\bibitem{Ellis:2013xoa}
J.~Ellis, D.~V. Nanopoulos  and K.~A. Olive, \emph{{No-Scale Supergravity
  Realization of the Starobinsky Model of Inflation}}, Phys. Rev. Lett. {\bf
  111} (2013) \href{http://dx.doi.org/10.1103/PhysRevLett.111.129902,
  10.1103/PhysRevLett.111.111301}{111301},
  \href{http://arxiv.org/abs/1305.1247}{{\tt arXiv:1305.1247 [hep-th]}},
[Erratum: Phys. Rev. Lett.111,no.12,129902(2013)]
%%CITATION = ARXIV:1305.1247;%%.

\bibitem{Farakos:2013cqa}
F.~Farakos, A.~Kehagias  and A.~Riotto, \emph{{On the Starobinsky model of
  inflation from supergravity}}, Nucl.Phys. {\bf B876} (2013)
  \href{http://dx.doi.org/10.1016/j.nuclphysb.2013.08.005}{187--200},
\href{http://arxiv.org/abs/1307.1137}{{\tt arXiv:1307.1137}}
%%CITATION = ARXIV:1307.1137;%%.

\bibitem{Ferrara:2013rsa}
S.~Ferrara, R.~Kallosh, A.~Linde  and M.~Porrati, \emph{{Minimal Supergravity
  Models of Inflation}}, Phys. Rev. {\bf D88} (2013), no.~8,
  \href{http://dx.doi.org/10.1103/PhysRevD.88.085038}{085038},
\href{http://arxiv.org/abs/1307.7696}{{\tt arXiv:1307.7696 [hep-th]}}
%%CITATION = ARXIV:1307.7696;%%.

\bibitem{Ferrara:2014kva}
S.~Ferrara, R.~Kallosh  and A.~Linde, \emph{{Cosmology with nilpotent
  superfields}}, JHEP {\bf 10} (2014)
  \href{http://dx.doi.org/10.1007/JHEP10(2014)143}{143},
\href{http://arxiv.org/abs/1408.4096}{{\tt arXiv:1408.4096 [hep-th]}}
%%CITATION = ARXIV:1408.4096;%%.

\bibitem{Ferrara:2015cwa}
S.~Ferrara and A.~Sagnotti, \emph{{Supersymmetry and inflation}}, in {\em {14th
  Marcel Grossmann Meeting on Recent Developments in Theoretical and
  Experimental General Relativity, Astrophysics, and Relativistic Field
  Theories (MG14) Rome, Italy, July 12-18, 2015}}.
\newblock 2015.
\newblock \href{http://arxiv.org/abs/1509.01500}{{\tt arXiv:1509.01500
  [hep-th]}}.
\newblock
\url{http://inspirehep.net/record/1391782/files/arXiv:1509.01500.pdf}.
\newblock
%%CITATION = ARXIV:1509.01500;%%.

\bibitem{Bergshoeff:2016psz}
E.~Bergshoeff, D.~Freedman, R.~Kallosh  and A.~Van~Proeyen, \emph{{Construction
  of the de Sitter supergravity}}, in {\em {About Various Kinds of
  Interactions: Workshop in honour of Professor Philippe Spindel Mons, Belgium,
  June 4-5, 2015, eds. N. Boulanger and S. Detournay, 2017}}.
\newblock 2016.
\newblock \href{http://arxiv.org/abs/1602.01678}{{\tt arXiv:1602.01678
  [hep-th]}}.
\newblock
\url{http://inspirehep.net/record/1419685/files/arXiv:1602.01678.pdf}.
\newblock
%%CITATION = ARXIV:1602.01678;%%.

\bibitem{Ferrara:2016ajl}
S.~Ferrara, A.~Kehagias  and A.~Sagnotti,
  \href{http://dx.doi.org/10.1142/S0217751X16300441}{\emph{{Cosmology and
  supergravity}},} in {\em {Memorial meeting for Nobel laureate Professor Abdus
  Salam's 90th birthday Singapore, January 25-28, 2016}}, vol.~A31, p.~1630044.
\newblock 2016.
\newblock
\href{http://arxiv.org/abs/1605.04791}{{\tt arXiv:1605.04791 [hep-th]}}.
\newblock
%%CITATION = ARXIV:1605.04791;%%.

\bibitem{Pestun:2016zxk}
V.~Pestun {\em et al.}, \emph{{Localization techniques in quantum field
  theories}},
\href{http://arxiv.org/abs/1608.02952}{{\tt arXiv:1608.02952 [hep-th]}}
%%CITATION = ARXIV:1608.02952;%%.

\bibitem{Pufu:2016zxm}
S.~S. Pufu, \emph{{The $F$-theorem and $F$-maximization}}, in {\em Localization
  techniques in quantum field theories (eds. V. Pestun and M. Zabzine)}.
\newblock 2016.
\newblock \href{http://arxiv.org/abs/1608.02960}{{\tt arXiv:1608.02960
  [hep-th]}}.
\newblock
\url{https://inspirehep.net/record/1480387/files/arXiv:1608.02960.pdf}.
\newblock
%%CITATION = ARXIV:1608.02960;%%.

\end{thebibliography}
\end{document}